\RequirePackage{fix-cm}
\documentclass[twocolumn]{svjour3}          
\smartqed  
\usepackage{graphicx}
\usepackage{subfigure}
\usepackage{multirow}
\usepackage{url}
\usepackage[utf8]{inputenc}
\usepackage{epstopdf}
\usepackage[inline]{enumitem}
\usepackage{mwe}
\usepackage[export]{adjustbox}
\usepackage{amsmath}
\usepackage{mathtools}
\usepackage{array}
\usepackage{booktabs}
\usepackage[nolist,nohyperlinks]{acronym}
\usepackage{cite}
\begin{document}

\title{Objective quality assessment of medical images and videos: Review and challenges}


\author{Rafael Rodrigues*,
		Lucie L\'{e}v\^{e}que,
            Jes\'us Guti\'{e}rrez,
            Houda Jebbari,
            Meriem Outtas,
            Lu Zhang,
            Aladine Chetouani,
            Shaymaa Al-Juboori,
            Maria Martini,
		Antonio~M.~G.~Pinheiro}

\institute{
\textbf{R.~Rodrigues and A.~M.~G.~Pinheiro}\newline Instituto de Telecomunicações (IT) and  Universidade da Beira Interior, Covilhã, PT \newline \textbf{L.~L\'{e}v\^{e}que} \newline Nantes Laboratory of Digital Sciences (LS2N), Nantes University, Nantes, FR \newline \textbf{J.~Guti\'errez} \newline Information Processing and Telecommunications Center and Universidad Polit\'ecnica de Madrid, Madrid, ES \newline \textbf{H.~Jebbari, M.~Outtas and L.~Zhang}\newline National Institute of Applied Sciences (INSA), Rennes, FR \newline \textbf{A.~Chetouani} \newline PRISME Laboratory, University of Orl\'eans,  Orl\'eans, FR \newline \textbf{S.~Al-Juboori and M. Martini} \newline Kingston University, London, UK \newline \newline Corresponding author: R.~Rodrigues \newline \textit{e-mail: rafael.rodrigues@ubi.pt}
}

\authorrunning{R. Rodrigues \textit{et al}} 


\maketitle

\begin{abstract}
Quality assessment is a key element for the evaluation of hardware and software involved in image and video acquisition, processing, and visualization. In the medical field, user-based quality assessment is still considered more reliable than objective methods, which allow the implementation of automated and more efficient solutions. Regardless of increasing research in this topic in the last decade, defining quality standards for medical content remains a non-trivial task, as the focus should be on the diagnostic value assessed from expert viewers rather than the perceived quality from na\"{i}ve viewers, and objective quality metrics should aim at estimating the first rather than the latter. In this paper, we present a survey of methodologies used for the objective quality assessment of medical images and videos, dividing them into visual quality-based and task-based approaches. Visual quality based methods compute a quality index directly from visual attributes, while task-based methods, being increasingly explored, measure the impact of quality impairments on the performance of a specific task. A discussion on the limitations of state-of-the-art research on this topic is also provided, along with future challenges to be addressed.


\end{abstract}

\section{Introduction}


\begin{figure*}[t!]
    \centering
    \includegraphics[width=0.8\textwidth]{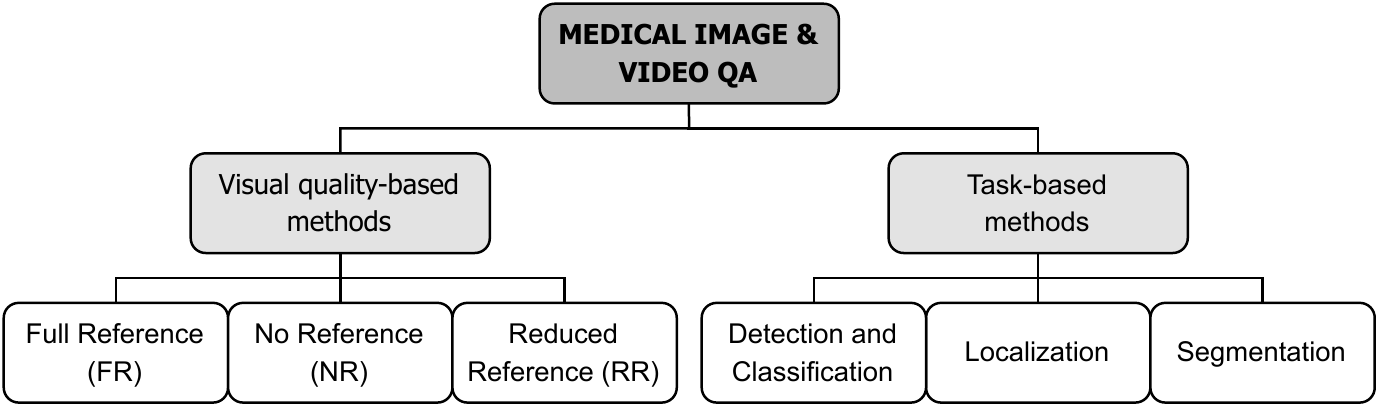}
    \caption{Overview of the paper organization (QA: quality assessment).}
    \label{fig:overview}
\end{figure*}

Medical images and videos provide clinical information from the human body with reduced invasiveness, but also structural and functional outcomes which could not be obtained by other means. Nowadays, medical imaging plays an inexorable part in diagnosis, treatment planning, and patient monitoring. Over the last decades, medical imaging techniques have been constantly developed, updated, and extensively used in numerous medical specialties. The World Health Organization (WHO) estimated around 3.6 billion diagnostic procedures performed worldwide, per year, between 1997 and 2007~\cite{world2016communicating}. Radiology is the leader in the production of medical imaging content~\cite{krupinski2010}, with a vast number of imaging modalities. Data from the \ac{OECD}, which considers \ac{CT}, \ac{MRI}, and \ac{PET} scans performed each year, further reveals a clear upward trend in the last decade, with over 253 million exams reported by \ac{OECD} countries~\cite{OECDhealth}. Moreover, and besides other relevant imaging modalities, medical images and videos are now also transmitted in real time for telemedicine applications. Therefore, large amounts of content from different acquisition methods are continuously created in the medical practice.

Several impairments can affect the quality of the end visual signal, impacting the quality perceived by the viewers (e.g., clinicians, radiologists, etc.), as well as their performance on clinical tasks. This spectrum of impairments highly depends on a wide variety of acquisition and reconstruction-related factors, which are often specific to each imaging modality. Furthermore, external and patient-related factors may induce artifacts in the acquired content, as well. Images and videos may also be subject to different processing, encoding/compression, transmission, and visualization methodologies. For example, in telemedicine applications, video content and real-time interactions are key features, which lead to a high demand of hardware resources and network bandwidth for data storage and transmission~\cite{leveque2017, chaabouni14}. Thus, measuring image and video quality in health applications is both a necessity, towards improving methodologies throughout the clinical workflow, and a wide open challenge, given the diversity of content, applications, and impairments. This paper aims at providing researchers with a broad sense of current boundaries and future opportunities in this field.

\section{Overview of Image and Video Quality Assessment}

The quality perceived by the users of image and video content is one of several factors influencing their \ac{QoE}, defined by EU Cost Action 1003 Qualinet~\cite{le2013qualinet} as ``the degree of delight or annoyance of the user of an application or service. It results from the fulfillment of his or her expectations concerning the utility and enjoyment of the application or service in the light of the user’s personality and current state''.

Typically, image and video quality is evaluated from a perceptual perspective, either subjectively or objectively, with each approach bearing its own motivations, advantages, and drawbacks. Subjective \ac{QA} relies on human observers to analyze images or videos, and rate their quality following specific methodologies (e.g.,~\cite{BT500, P913}). The output of subjective tests is then statistically analyzed, usually by computing the \ac{MOS} or \ac{DMOS}~\cite{nasr2017visual}. Depending on the type of content and application, user expertise may be a requirement for an accurate and meaningful evaluation. In most healthcare applications, the quality perception of end-users is likely to be strongly influenced by the clinical utility of the content, rather than strictly aesthetic criteria. Therefore, subjective \ac{QA} methods may have some limitations in this context, concerning the availability of expert observers, in addition to the inherent intra- and inter-subject scoring variability~\cite{mason2019comparison}.

While a recent review of subjective \ac{QA} of medical images and videos was published by Lévêque \textit{et al.} \cite{leveque2021}, our paper focuses on objective \ac{QA} methods for medical images and videos. Objective \ac{QA} relies on the use of image processing and analysis algorithms towards automated quality estimation methods~\cite{nasr2017visual}. Although objective methods may be less reliable than subjective assessment, they are more cost-effective, less time-consuming, and  reduce observer variability and bias~\cite{mason2019comparison}. In another review paper published in 2016, Chow \textit{et al.} discussed both subjective and objective methods~\cite{chow2016review}, with a primary focus on \ac{MRI}, \ac{CT}, and ultrasound imaging. More recently, Raj \textit{et al.} \cite{raj2019fundus} presented a survey of objective \ac{QA} methods for fundus images. In our article, we will mainly focus on more recent research, and on imaging modalities not included in these last two papers.

The most common design principle in objective image and video \ac{QA} applications is the computation of a quality index from certain visual and/or structural attributes of the content. On the contrary, task-based \ac{QA} methods assess the quality in terms of a specific goal of the content, by measuring its influence on the performance of certain tasks (e.g., diagnosis or localization of an anatomical structure~\cite{barrett2015task}). This kind of approach has been increasingly explored in the development of quality metrics that are specific for medical images and video. Consequently, we propose to divide the discussed \ac{QA} methodologies into two major categories, i.e., visual quality-based, and task-based methods, as described in Figure \ref{fig:overview}.

\section{Visual Quality-based Methods}

Visual quality-based methods are commonly categorized as \ac{FR}, \ac{NR}, or \ac{RR}, depending on the availability of an undistorted reference (preferably the original image or video). \ac{FR} metrics predict the quality by directly comparing the reference and its distorted versions, contrary to \ac{NR} metrics which assess the quality only using the distorted image. Finally, \ac{RR} metrics compare features representative of the distorted and reference images.
The performance of objective methods is usually evaluated through a statistical comparison with subjective results~\cite{BT500,ITU-T2012}.
%
The following subsections present an overview of visual quality-based objective \ac{QA} methods for medical images and video, according to the aforementioned categorization.

\subsection{Full-Reference Approaches}

\begin{table*}[t!]
\renewcommand{\arraystretch}{1.4}
    \centering
    \small
    \caption{Overview of visual quality-based approaches to the quality assessment of medical images and video (FR metrics). In some works, NR metrics were also tested, which are marked with $^{*}$.}
    \vspace{0.2cm}
    \begin{tabular}{>{\raggedright}m{0.09\textwidth} >{\raggedright}m{0.15\textwidth} >{\raggedright}m{0.21\textwidth} >{\raggedright}m{0.18\textwidth} >{\raggedright}m{0.245\textwidth}}
    
        \toprule
        \textbf{Reference} 	& \textbf{Imaging modality} & \textbf{Quality impairments / Processing} 	& \textbf{Validation}	& \textbf{Objective metrics}
        \tabularnewline
        \midrule
       
        \cite{chow2016correlation}		   & MRI	& Rician and Gaussian noise, and Gaussian blur (5 levels); DCT, JPEG and JPEG2000 compression (5 ratios)
        & MOS: 4 medical observers (SDSCE)     & SNR, PSNR, SSIM, MSSIM, FSIM, ICF, NQM, WSNR, VIF, VIFP, UQI, IW-PSNR,  IW-SSIM
        \tabularnewline %
        
        \cite{mason2019comparison}		   & MRI	& White and Rician noise, Gaussian blur, motion, undersampling and wavelet compression		
        & MOS: 5 radiologists    & RMSE, PSNR, SSIM, MS-SSIM, IW-SSIM, GMSD, FSIM, HDRVDP, NQM, VIF
        \tabularnewline
        
        \cite{zhou2003practice} & CT & Compression with DCT and Wavelet using different compression ratios & MOS: Common persons, clinic doctors, and radiologists & PSNR, SSIM, distortion measure from \cite{franti1998blockwise}.
        \tabularnewline
        
        
		\cite{kowalik2015modelling}		   & CT	& JPEG and JPEG2000 compression (5 ratios)		& MOS: 6 radiologists (Double-stimulus DCR)		& MSE, local MSE, SNR, SSIM, VSNR, VIF
		\tabularnewline
        
		\cite{panayides2011atherosclerotic} & Ultrasound videos & H.264 Compression with different quantization parameters and packet-loss rates & MOS: 2 medical experts & PSNR, SSIM, VSNR, VIF, VIFP, IFC, NQM, WSNR
		\tabularnewline
		
		\cite{razaak2014study}	   	&  Ultrasound videos			& HEVC compression (8 quantization levels)		& MOS: 4 medical experts and 16 na\"ive observers (DSCQS) 		& PSNR, SSIM, UQI, VQM, NQM, VIF, VSNR
		\tabularnewline
		
		\cite{razaak2016cuqi}	   	&  Ultrasound videos								& HEVC compression (8 quantization levels)		& DMOS: 4 medical experts (DSCQS) 		& PSNR, SSIM, UQI, VQM, NQM, VIF, VSNR, CUQI (proposed metric)
		\tabularnewline
		
		\cite{kumcu2014visual}	   	& Laparoscopic surgery videos								& H.264 compression (4 bit rates)		& MOS: 9 laparoscopic surgeons and 16 na\"ive observers (SSCQE) 		& VQM, HDR-VDP-2, PSNR
        \tabularnewline
		
        \cite{chaabouni14}		   	& Endoscopic surgery videos		& H.264 compression (11 ratios)		& MOS: 14 medical observers (DSCQS) 		& SSIM, UQI, PSNR, WSNR, VSNR, HDR-VDP, IFC, MSE, MS-SSIM, PSNR-HVS, PSNR-HVS-M, VIF, VIFP, NIQE$^{*}$ and BRISQUE$^{*}$
        \tabularnewline
		
	    \cite{usman2017quality}	   	& Endoscopic videos								& HEVC compression (8 quantization levels)		& MOS: 6 medical observers and 19 na\"ive observers (DSCQS) 		& MSE, PSNR, SSIM, MS-SSIM, VSNR, IFC, VIF, VIFP, UQI, NQM, WSNR
	    \tabularnewline
		
	    \cite{kumar2009mos}	   	& CT, MRI &  Varying bitrates (from 0.1 to 2.0) using SPIHT compression & MOS: 6 medical observers & 	PSNR, SSIM
	    \tabularnewline
		
		\cite{kumar2013development} & CT, MRI, Ultrasound &  Varying bitrates (from 0.02 to 2.0) using SPIHT compression & MOS: 5 medical observers & 	PSNR, SSIM, proposed SSIM variation 
		\tabularnewline
		
		
		\cite{renieblas2017structural}		   	  & X-ray, MRI			& Gaussian noise and Gaussian blur (5 levels); JPEG and JPEG2000 compression (5 ratios)		& MOS: 3 radiologists (Double-stimulus DCR) 		& 16 combinations of SSIM components (multiscale, gradient-based and structural component)
		\tabularnewline
		
		\bottomrule
    \end{tabular}
    \label{tab:vq_based_fr}
\end{table*}

Full-reference QA of medical images and videos 
had
focused almost exclusively on adapting metrics designed for non-medical content to obtain quality predictions for compressed and/or artificially distorted medical content. Table \ref{tab:vq_based_fr} summarizes the works presented in this section.

\subsubsection{Magnetic Resonance Imaging}
\ac{FR} quality assessment studies using \ac{MRI} 
includes
the works of Chow \textit{et al.}~\cite{chow2016correlation} and Mason \textit{et al.}~\cite{mason2019comparison}. Both works tested a large set of metrics, e.g. \cite{ssim, fsim, ifc, nqm, vif, uqi} on datasets with different compression settings (\ac{DCT}~\cite{dct}, JPEG~\cite{jpeg}, JPEG2000~\cite{jpeg2000}, 
wavelets) and simulated artifacts (e.g. Rician noise, Gaussian blur or motion artifacts). The best average correlations with the subjective scores were achieved 
for
\ac{NQM} 
with
\ac{PLCC} and \ac{SROCC} around 0.94
\cite{chow2016correlation}, and \ac{VIF} 
\cite{mason2019comparison}.


\subsubsection{Computed Tomography}

In an early work from 2003, Zhou \textit{et al.} \cite{zhou2003practice} estimated the quality of \ac{CT} images with both \ac{DCT} and Wavelet compression, using a back-propagation neural network with \ac{PSNR} and \ac{SSIM}, and the Block-wise Distortion Measure \cite{franti1998blockwise} as inputs. The reported agreement rates with the obtained subjective scores ranged between 85.71\% and 96.88\%. Kowalik \textit{et al.}\cite{kowalik2015modelling} also used \ac{CT} images with compression - JPEG and JPEG2000, and studied the \ac{ROC} curves of \ac{SSIM}, \ac{SNR}, \ac{VIF}, \ac{MSE}, and \ac{VSNR}~\cite{vsnr} in classifying images as having non-noticeable or non-acceptable distortions, after subjective annotation by experts. The largest \ac{AUC} was obtained with \ac{SSIM} (0.99 for brain images and 0.96 for body images), closely followed by \ac{VIF}.

\subsubsection{Ultrasonography}

In \cite{panayides2011atherosclerotic}, the authors evaluated the performance of a H.264 encoding framework for atherosclerotic plaque ultrasound videos, which resulted in enhanced performance in noisy environments. WSNR obtained the best correlation with the subjective scores (\ac{PLCC} = 0.69 and \ac{SROCC} = 0.72). The authors also proposed minimum settings for several parameters, including the frame rate, bit rate, and PSNR in the \ac{ROI}.

The authors of \cite{razaak2014study} and \cite{razaak2016cuqi} assessed the quality of ultrasound video excerpts compressed with \ac{HEVC}~\cite{hevc}, using a set of 7 common \ac{FR} metrics, as well as the proposed content-specific \ac{CUQI}, in \cite{razaak2016cuqi}. \ac{CUQI} estimates the diagnostic quality of cardiac ultrasound video, using motion and edge information. In terms of correlation with \ac{DMOS}, all metrics yielded good \ac{PLCC} and \ac{SROCC} coefficients in both studies. In \cite{razaak2016cuqi}, \ac{CUQI} outperformed the other tested metrics, with \ac{PLCC} and \ac{SROCC} after \ac{DMOS} nonlinear regression of 0.94 and 0.93, respectively. Nonetheless, all the tested metrics achieved correlations above 0.9, and \ac{SSIM} achieved the best result in terms of \ac{PLCC} without nonlinear regression of \ac{DMOS}.

\subsubsection{Endoscopic / Laparoscopic videos}

In \cite{kumcu2014visual} and \cite{chaabouni14}, the authors performed objective \ac{QA} of laparoscopic and surgical videos, respectively, compressed with H.264. \ac{SSIM}, \ac{HDR-VDP-2}, and \ac{VQM} were used in both studies, but a larger set of metrics was used in \cite{chaabouni14}, including, for example, PSNR-HVS and PSNR-HVS-M~\cite{psnr-hvs}, \ac{MSE},\ac{MS-SSIM}~\cite{ms-ssim}, and two \ac{NR} metrics - \ac{NIQE}~\cite{niqe} and \ac{BRISQUE}~\cite{brisque}. In the latter study, \ac{MSE}, \ac{SSIM}, \ac{MS-SSIM}, and \ac{BRISQUE} achieved good correlations with \ac{MOS} (\ac{PLCC} $>$ 0.9).

Usman \textit{et al.}\cite{usman2017quality} were the first to investigate the quality of wireless capsule endoscopy videos. Videos were compressed with \ac{HEVC}, and 10 metrics were computed to estimate video quality. \ac{VIF}, \ac{VIFP}, and \ac{IFC} showed good correlation with the obtained \ac{MOS} (between 0.88 and 0.92), with \ac{VIFP} outperforming the other two, both in statistical significance and in computation time.

\subsubsection{Multiple imaging modalities}

Other interesting works were published in \cite{kumar2009mos} and \cite{kumar2013development}, where the authors studied the performance of \ac{PSNR} and \ac{SSIM} on \ac{CT}, \ac{MRI}, and ultrasound images with \ac{SPIHT} compression~\cite{spiht}. Both \ac{PSNR} and \ac{SSIM} quality estimations correlated well with subjective scores. In \cite{kumar2013development}, Kumar \textit{et al.} proposed an improved version of \ac{SSIM}, which outperformed \ac{PSNR} with correlation coefficients of 0.99 (\ac{MRI}), 0.98 (\ac{CT}), and 0.98 (ultrasound).

Several variations of \ac{SSIM} were also tested by Renieblas \textit{et al.}\cite{renieblas2017structural}, to estimate the quality of planar X-ray and \ac{MRI}, with simulated Gaussian blur and noise, and JPEG and JPEG2000 compression. Best overall correlation with subjective scores were obtained with 4-MS-G-SSIM (PLCC $\geq$ 0.75 overall, and $\geq$ 0.86 for MR images), which considers four-component region-based weighting (4-), multiscale (MS-), and gradient-based (G-) computation. The authors concluded that MS- approaches generally improve the performance of single-scale counterparts, and that $r*$ (structural component only) showed a slight advantage over the complete \ac{SSIM} index.

\subsection{{Reduced-Reference Approaches}}

A few works, described in Table \ref{tab:vq_based_rr}, addressed the use of \ac{RR} metrics to measure the quality of medical images and video. One of the most popular is the \ac{RR} version of \ac{VQM}, which, as aforementioned, was used in \cite{kumcu2014visual}, showing a reasonable performance (\ac{PLCC}=0.97 and \ac{SROCC}=0.94) in comparison to other \ac{FR} metrics. Other approaches are based on the comparison of certain attributes of the images or videos, through the computation of a similarity score. For example, Lee and Wang used the similarity between the intensity histograms of reference and distorted fundus images to provide an estimation of their quality~\cite{lee1999automatic}. For the same image modality, the authors in \cite{lalonde2001automatic} proposed a similarity metric based on the comparison of the distribution of edge magnitudes and the local intensity distribution for distorted and reference images. In both cases, the performance of their approaches was not reported with objective measures, but through a comparison with SNR and through qualitative results, respectively. The results show that the proposed approaches are useful to discriminate between good and bad images. In addition, the proposed approaches can be extended to other types of images. 

\begin{table*}[t!]
\renewcommand{\arraystretch}{1.4}
    \centering
    \small
    \caption{Overview of visual quality-based approaches to the quality assessment of medical images and video (RR metrics).}
    \vspace{0.2cm}
    \begin{tabular}{>{\raggedright}m{0.09\textwidth} >{\raggedright}m{0.15\textwidth} >{\raggedright}m{0.21\textwidth} >{\raggedright}m{0.18\textwidth} >{\raggedright}m{0.245\textwidth}}
    
        \toprule
        \textbf{Reference} 	& \textbf{Imaging modality} & \textbf{Quality impairments / Processing} 	& \textbf{Validation}	& \textbf{Objective metrics}
        \tabularnewline
        \midrule
        
		\cite{lee1999automatic}	   	& Retinal fundus photography		& Reduction of brightness, contrast, and SNR &
		Comparison with SNR and agreement with human perception  & Similarity metric based on intensity histograms (RR)
		\tabularnewline
		
		\cite{lalonde2001automatic}	   	& Retinal fundus photography	& Image acquisition impairments (out of focus, poor illumination, etc.) & Qualitative assessment: 1 ophthalmologist  & Similarity metric based on the distribution of edge magnitudes and the local intensity distribution (RR)
		\tabularnewline
		
		\cite{planitz2005study}	   	& CT, MRI & Compression  with  DCT  and  Least Significant Bit (LSB) techniques and different compression ratios & Objective measures of classification and percentage of damages pixels  & Structural comparison metric (RR).
		\tabularnewline
		
		\cite{nasr2017visual}	   	& Ultrasound videos		& Gaussian noise, JPEG2000 compression (9 ratios), HEVC compression (8 QP) & DMOS: 4 medical experts, 16 naive observers (DSCQS) & PSNR, SSIM,  MSSIM 
		\tabularnewline
		
		\bottomrule
    \end{tabular}
    \label{tab:vq_based_rr}
\end{table*}

Finally, another set of RR metrics is based on the use of watermarking, i.e., adding a payload to the content, and then using similarity metrics to compare the reference and the distorted image or video. An example of this approach was proposed by Planitz and Maeder~\cite{planitz2005study}, using \ac{SSIM}to measure the degradations and watermarking capacity. This study demonstrated that more robust watermarking techniques can be used in less visually sensitive areas, while lighter techniques should be applied in 
more sensitive areas. Nasr and Martini\cite{nasr2017visual} used \ac{PSNR} and \ac{MSSIM} to measure the quality of medical ultrasound videos with simulated Gaussian, and JPEG2000 and \ac{HEVC} compression. A predefined reduced-size logo that shares the same features of the original frame of interest (i.e., same organ and layout). Considering that the logo does not depend on the specific content of the sequence, this method can be considered, to some extent, as \ac{NR}. The authors concluded that the proposed technique did not require the original frame, whilst achieving a high \ac{PLCC} with the subjective results, with reported average correlations above 0.97, for both PSNR and MSSIM.

\begin{table*}[t!]
\renewcommand{\arraystretch}{1.4}
    \centering
    \small
    \caption{Overview of visual quality-based approaches to the quality assessment of medical images and video (NR metrics I).}
    \vspace{0.2cm}
    \begin{tabular}{>{\raggedright}m{0.09\textwidth} >{\raggedright}m{0.15\textwidth} >{\raggedright}m{0.21\textwidth} >{\raggedright}m{0.18\textwidth} >{\raggedright}m{0.245\textwidth}}
    
        \toprule
        \textbf{Reference} 	& \textbf{Imaging modality} & \textbf{Quality impairments / Processing} 	& \textbf{Validation}	& \textbf{Objective metrics}
        \tabularnewline
        \midrule
        
        \cite{Liebgott16ICASSP}		   	& MRI		& Different imaging sequences, contrast weights and subsampling strategies	& 5-level qualitative scale: 5 observers & Contrast, resolution, texture and intensity-based features with SVM classifier
        \tabularnewline
        
        \cite{chow2017modified}		   	& MRI		& Rician and Gaussian noise, and Gaussian blur (5 levels); DCT, JPEG and JPEG2000 compression (5 ratios)		& MOS: 10 radiologists; \newline FR metrics: NQM and FSIM		& BRISQUE (modified and original), JPEG-based model
        \tabularnewline
        
		\cite{osadebey2017bayesian}		   	& MRI							& Bias fields, motion blur, Rician noise (20 levels)		& MOS: 4 radiologists and 1 experienced MRI reader		& Entropy posterior probability
		\tabularnewline
		
		\cite{obuchowicz2020magnetic}		   	& MRI										& Not specified		& MOS: 31 radiologists 		& ENMIQA (proposed entropy-based metric), BPRI, dipIQ, IL-NIQE, MEON, $Q$ index, $S$ index, QENI, SNR-ROI, SSEQ and SISBLIM
		\tabularnewline
		
		 \cite{chabert2021image}		   	& MRI										& Gaussian noise, Gaussian blurring, contrast manipulation 		& MOS: 3 neuroradiologists 		& Whole image and ROI-based features (e.g. SNR, CNR, uniformity, Wang index, Shannon entropy) with different classifiers (LDA, QDA, SVM, MLP, Logistic Regression)
        \tabularnewline
		
		\cite{esses2018automated}		   	& MRI		& Not specified (motion artifacts and inhomogeneous fat supression mentioned in some cases)		& Binary quality label: 2 radiologists & AlexNet
		\tabularnewline
		
		\cite{sujit2019automated}		   	& MRI										& Motion artifacts	& Qualitative binary score: 2 expert observers 	& 6-layer CNN 
        \tabularnewline
		
		\cite{ma2020diagnostic}		   	& MRI										& Motion artifacts		& Qualitative binary score and 3-level scales: 2 radiologists 		& 4-layer CNN and ResNet-10
		\tabularnewline
    
        
        \cite{kohler2013automatic}	& Retinal fundus \newline photography       	& Gaussian blurring        (2 levels) and Gaussian noise (20 levels)		& FR metrics: PSNR and SSIM; MOS (no. of observers not given)		& Local gradient-based metrics (normal and vessel-guided), CPBD and anisotropy
        \tabularnewline
        
        \cite{wang2015human}	& Retinal fundus \newline photography       	& Uneven illumination, blurring, low contrast and color distortion		& Composite binary scoring: 3 ophthalmologists		& HVS-based feature matrix (multiple channel sensation, JNB, contrast sensitivity function)
        \tabularnewline
        
        \cite{remeseiro2017objective}	& Retinal fundus \newline photography       	& Not specified		& MOS (no. of observers not given)		& Texture features from RGB and CIELab channels (Butterworth, Gabor and Wavelet filters, Gaussian Markov random fields, GLCM)
        \tabularnewline

		\bottomrule
    \end{tabular}
    \label{tab:vq_based_nr1}
\end{table*}

\subsection{No-Reference Approaches}

Tables \ref{tab:vq_based_nr1} and \ref{tab:vq_based_nr2} summarize the articles further presented in this section. 

\subsubsection{Magnetic Resonance Imaging}

Liebgott \textit{et al.}\cite{Liebgott16ICASSP} proposed a new method to assess the quality of 2D \ac{MRI} using active learning. A descriptor including features based on contrast, resolution, texture, and intensity information was reduced using \ac{PCA}~\cite{pca} and classified using an \ac{SVM} model~\cite{libsvm}. Subjetive socres on a 5-level scale were used to train and validate the model. Results showed that active learning reduces the need for training data by around 50\%, compared to a previous method by the same authors.

Chow and Rajagopal proposed a modified version of \ac{BRISQUE}\cite{chow2017modified}, trained with mean-subtracted contrast-normalized \ac{MRI} and the corresponding \ac{DMOS}. This proposed metric was evaluated on two separate datasets with both T1-weighted (T1\emph{w}) and T2\emph{w} brain \ac{MRI}, one with unknown artifacts, and another with the same range of distortions of \cite{chow2016correlation}. The proposed modified \ac{BRISQUE}, the original counterpart, and a JPEG-based model were tested on the first dataset, and the correlation with \ac{MOS} was obtained with the proposed metric (\ac{PLCC} of 0.96 and \ac{SROCC} of 0.93). The proposed \ac{BRISQUE} also outperformed its original version, in terms of correlation with two \ac{FR} metrics - \ac{NQM} and \ac{FSIM} -, for the majority of the simulated distortions in the second dataset.

The authors in \cite{osadebey2017bayesian} applied the Bayes theorem to calculate the posterior probability of entropy, given three quality attributes, i.e., contrast, sharpness, and standard deviation. A global quality index for 2D \ac{MRI} was obtained by averaging those probability values, which were first separately computed for low- and high-entropy feature images. The evaluation dataset included ten T1\emph{w} MRI of the brain acquired with bias fields, and twenty-one images acquired without perceived distortions (T1\emph{w}, T2\emph{w}, Proton Density and \ac{FLAIR}). Twenty different levels of Rician noise and motion blur were induced in the second subset. Five radiologists performed subjective \ac{QA}, which showed that the predicted quality decreased consistently across the twenty distortion levels of noise and blurring. Correlation with \ac{MOS} was assessed separately for each modality and distortion level using \ac{SROCC}. The reported coefficients were globally above 0.6, with a tendency to be higher for lower distortion levels.

In \cite{obuchowicz2020magnetic}, the authors presented ENMIQA, an entropy-based metric for the objective \ac{QA} of \ac{MRI}, which expresses local intensity differences after non-maximum suppression at various threshold levels. A dataset of T2\emph{w} \ac{MRI} was used, but no information was provided in the paper on the type of artifacts or noise present in the images. The performance of the proposed metric in quality prediction was compared against a large set of \ac{NR}, and ENMIQA outperformed them all in terms of correlation with \ac{MOS}. However, overall correlation coefficients were quite low, especially for rank correlations (\ac{PLCC} = 0.65, \ac{SROCC} = 0.35). \ac{PLCC} was also reported for each anatomical structure separately, with wrist and knee images yielding coefficients near 1 and 0.9, respectively.

In a recent work, Chabert \textit{et al.}\cite{chabert2021image} proposed using a set of features reduced using \ac{PCA}, which included \ac{SNR}, \ac{CNR}, \ac{FBER}, sharpness in fat, uniformity, the Wang index~\cite{wang2002no} and Shannon entropy~\cite{shannonE}, among others, to predict the quality perception of neuroradiologists for lumbar T1\emph{w} and T2\emph{w} \ac{MRI}, with simulated Gaussian noise and blurring, and contrast manipulation. A few classifiers were tested, with \ac{SVM} showing a superior overall performance (accuracy above 73\%). Despite relying on user interaction for \ac{ROI}-based feature extraction, the results seem promising for online monitoring of the image quality, at the moment of acquisition.

The appearance of deep learning models has been establishing a new paradigm in \ac{NR} medical image quality assesment. Esses \textit{et al.}\cite{esses2018automated} trained a \ac{CNN} model based on the AlexNet architecture~\cite{alexnet} to screen the diagnostic quality of T2\emph{w} liver \ac{MRI}. The obtained results indicate a 79\% and 73\% concordance between \ac{CNN} predictions and experts 1 and 2, respectively. Moreover, the \ac{CNN} showed good negative predictive values in the identification of non-diagnostic image quality (94\% and 86\% for each of the experts).

Sujit \textit{et al}.~implemented a branched custom \ac{CNN} model to assess the image quality of 3D T1\emph{w} brain \ac{MRI}~\cite{sujit2019automated}.~Each branch received a different imaging plane (i.e., coronal, sagittal, and axial) as input. The global quality score was obtained by averaging the scores of each plane. Training and validation data were collected from the Autism Brain Imaging Data Exchange (ABIDE) multicentric dataset~\cite{di2014autism}, which is publicly available with binary subjective quality scores (i.e., "acceptable" or "unacceptable"). The proposed ensemble model obtained an \ac{AUC} of 0.90 on a 20\% test split. The model was also tested on a separate test set, from the CombiRx dataset~\cite{lublin2013randomized}, yielding a lower \ac{AUC} of 0.71. The authors argued that the lower results may be due to differences in cohorts, as CombiRx data was not included in training.

Ma \textit{et al.}\cite{ma2020diagnostic} investigated the use of \ac{CNN} to estimate the diagnostic quality of abdominal \ac{MRI}. A custom 4-layer ac{CNN} and a ResNet-10 model, were trained to classify images on both binary (non-diagnostic vs. diagnostic) and 3-level (non-diagnostic (0), diagnostic (1), or excellent (2)) quality scales. The proposed 4-layer \ac{CNN} outperformed ResNet-10, with an accuracy of 84\% in binary classification (\ac{AUC} = 0.72) and 65\% in 3-label classification (AUC\textsubscript{0} = 0.77, AUC\textsubscript{1} = 0.69, AUC\textsubscript{2} = 0.83). However, the authors argued that the high disagreement between human observers, as well as the probability of label unreliability, could influence the results. Moreover, the activation maps suggest that low-level features have higher discriminative power, whereas deeper features do not have a great impact on these classification tasks, i.e., deeper models may not necessarily improve the results in similar applications.

\subsubsection{Retinal fundus images / Ophthalmology images}

K{\"o}hler \textit{et al.}\cite{kohler2013automatic} extended the work of \cite{zhu2010automatic}, where a global quality index $Q$ was obtained from patch-wise quality indices $q(P)$. These were based in the singular value decompositions of $G$, a patch-wise gradient matrix. In \cite{kohler2013automatic}, the authors implemented a spatially weighted version ($Q_v$), assigning larger weights to patches around blood vessels, according to a vesselness measure. The proposed metric was tested as a \ac{PSNR} and \ac{SSIM}, on images from the DRIVE database, with simulated Gaussian blur and noise. $Q_v$ outperformed $Q$, with an overall \ac{SROCC} of 0.89 (\ac{PSNR}) and 0.91 (\ac{SSIM}). In a second experiment, the authors performed binary quality prediction (acceptable vs. non-acceptable), considering annotations from experts. $Q_v$ outperformed other \ac{NR} metrics - $Q$, \ac{CPBD}, and an anisotropy measure -, with an \ac{AUC} of 0.89.


The authors in \cite{wang2015human} proposed an \ac{HVS}-based feature extraction algorithm, which relies on multi-channel sensation, \ac{JNB}, and the Contrast Sensitivity Function, to detect illumination and color distortions, blur, and low contrast in retinal fundus images. Extracted features were classified with an \ac{SVM} and a decision tree, to predict image quality, either based on each of the three aforementioned properties, or globally. A joint dataset contained the 3 partial binary annotations, given by ophthalmologists. Average \ac{AUC} of 0.97, 0.96, and 0.68 for illumination/color, \ac{JNB}, and color, respectively, was obtained with the decision tree, and 0.93, 0.93, and 0.88 with the \ac{SVM}. In terms of overall quality, the \ac{SVM} performed best, with sensitity/specificity of 0.92/0.87.

In \cite{remeseiro2017objective}, the authors used texture features from RGB and CIELab channels, including Wavelet and Gabor filters, and features from the \ac{GLCM}~\cite{glcm} to compute the quality of retinal fundus images. Several feature selection methods and classifiers were tested for the binary classification of the image quality (good / poor quality). However, no information on the quality impairments nor the subjective evaluation setup is provided in the paper. The best performance in quality prediction was obtained with \ac{GLCM} features from CIELab channels, filtered using correlation-based feature selection, and classified with an \ac{SVM} classifier (Accuracy = 99.09\%).

More recently, Coyner \textit{et al.}\cite{coyner2019automated} studied the use of \ac{CNN} models to measure the quality of retinal fundus images, considering their usefulness for a confident detection of retinopathy of prematurity. The model was based on the InceptionV3 architecture, and was trained to classify an independent set of thirty images as having acceptable or non-acceptable quality. The testing set was previously ranked by six experts, and the \ac{CNN} probability output showed good correlation with the consensus ranking from the experts (\ac{SROCC} = 0.90). \ac{SROCC} ranged from 0.86 to 0.93, when considering individual rankings. 

Focusing on eye fundus images for the diagnosis of diabetic retinopathy, Raj \textit{et al.}\cite{raj2020multivariate} presented a new multivariate regression-based \ac{CNN} model to predict the image diagnostic quality. The proposed model incorporates four different backbones - InceptionV3, DenseNet-121, ResNet-101, and Xception -, trained against subjective scores for six quality parameters, i.e., visibility of the optic disc, macula, and blood vessels, color, contrast, and blur. The top of the model provides a global diagnostic quality score based on the computed features. Results showed a strong correlation with the subjective scores for overall diagnostic quality, with values 0.94, 0.95, 0.85, and 0.40 obtained for \ac{SROCC}, \ac{PLCC}, \ac{KRCC}, and \ac{RMSE} respectively. The classification accuracy was 95.66\% over the FIQuA dataset, presented in the same paper, and 98.96\% and 88.43\% respectively over the two publicly available datasets DRIMDB~\cite{drimdb} and EyeQ~\cite{eyeQ}.

In \cite{shen2020domain}, the authors proposed a \ac{CNN}-based model with three modules for the quality assessment of retinal images. The model extracts global and local features (optic disc and fovea) using a VGG-16 backbone, which provides an overall quality score and three partial scores for different factors - artifacts, clarity and field definition. Visual feedback from class activation mapping is also provided to ophthalmologists. \ac{ROI} detection is performed by a separate module, using a ResNet-50 backbone for center detection and a VGG-16 local encoder for iterative \ac{ROI} refinement. A third module implements an unsupervised \ac{ADDA} method~\cite{adda}, using a Generative Adversarial Network architecture~\cite{gan} to address domain shifts between training and test data. Several variations of the proposed model were tested to classify images as adequate or inadequate for the diagnosis of diabetic retinopathy, and the best result was obtained with the full described model, with an \ac{AUC} of 0.95.

Niwas \textit{et al.}\cite{NIWAS16CMPB} developed a metric to assess the quality of \ac{AS-OCT} images using \ac{LBP} in the complex Wavelet domain. The image is first decomposed into 32 wavelet decomposition levels, using the Double Density Dual Tree-complex Wavelet transform, and the \ac{LBP} histogram is obtained for each sub-band. The Minimum Redundancy Maximum Relevance (mRMR) method is then applied to select the most relevant features to classify the image in terms of quality, with three levels considered. Several classifiers were tested, including \ac{SVM}, random forest, decision tree, and an AdaBoost classifier, but the best results were achieved using a Na{\"i}ve Bayes classifier with mRMR feature selection. These were compared to the ground truth given by experts, yielding an overall weighted accuracy of 82.9\%.

\begin{table*}[t!]
\renewcommand{\arraystretch}{1.4}
    \centering
    \small
    \caption{Overview of visual quality-based approaches to the quality assessment of medical images and video (NR metrics II).}
    \vspace{0.2cm}
    \begin{tabular}{>{\raggedright}m{0.09\textwidth} >{\raggedright}m{0.15\textwidth} >{\raggedright}m{0.21\textwidth} >{\raggedright}m{0.18\textwidth} >{\raggedright}m{0.245\textwidth}}
    
        \toprule
        \textbf{Reference} 	& \textbf{Imaging modality} & \textbf{Quality impairments / Processing} 	& \textbf{Validation}	& \textbf{Objective metrics}
        \tabularnewline
        \midrule
        
        \cite{coyner2019automated}	& Retinal fundus \newline photography       	& Not specified		& MOS: 3 ophthalmology experts (cross-validation set) + 6 ophthalmology experts (test set)		& InceptionV3
        \tabularnewline
        
        \cite{raj2020multivariate}		   	& Retinal fundus \newline photography 		& Not specified 		& Qualitative 3-level scale from feature-based MOS: 15 ophthalmologists	&  Multivariate regression-based CNN, with feature extraction using InceptionV3, DenseNet-121, ResNet-101, and Xception
        \tabularnewline
        
        \cite{shen2020domain}	& Retinal fundus \newline photography       	& Artifacts (e.g. shadows, lens stains), image clarity and field definition		&  Binary quality label and factor-specific qualitative scores: 3 ophthalmologists		& CNN model with global and local feature extraction and ADDA, DeepIQA, DRIQC, MEON, MFIQA
        \tabularnewline
        
        \cite{NIWAS16CMPB}		   	& AS-OCT		& Not specified 		& Qualitative 3-level scale: 2 medical experts	&  DDDT-CWT based LBP features with a na\"ive Bayes classifier
        \tabularnewline
        
        \cite{khan2020residual}	   	& Laparoscopic \newline surgery videos								& Defocus and motion blur, Gaussian noise, uneven illumination, smoke		& Labels inferred from simulated distortions 		& ResNet (18, 34, 50 layers), RankIQA
        \tabularnewline
        
        \cite{ali2021deep}	& Endoscopy video     	& Blur, bubbles, contrast, specularity, saturation, misc. artifacts		& MOS: 3 radiologists 	& YOLOv3 with DeepLabv3+ spatial pyramid pooling
        \tabularnewline

        \cite{abdi2017automatic}		   	& Ultrasound \newline (Echocardiogram)		& Not specified		& MES (6-level scale): \newline 1 cardiologist	& PSO optimized CNN model: 3 convolutional layers
        \tabularnewline
        
        \cite{tang2018no}	   	& Fused images (MRI, CT, US, SPECT, PET)		& 8 image fusion algorithms: NNM, LP-SR, CSCS, GFF, NSCT-PCNN-SF, ISML, CSR, and DTM-PCNN		& MOS: 20 radiologists 		& Pooling of phase congruency and standard deviation	(proposed), Gradient-based and Structure-based metrics, Edge information, RSFE, MI
        \tabularnewline
        
        \cite{TANG2020SPIC}	   	& Fused images (MRI, CT, US, SPECT, PET)		& 8 image fusion algorithms~\cite{tang2018no}		& MOS: 20 radiologists 		& PCNN in NSCT (proposed), baseline metrics of \cite{tang2018no}, Phase congruency, Entropy, OSSI, Tsallis entropy-based metric, BMPRI, MEON
        \tabularnewline
        
        \cite{outtas.study.16}		   	& MRI, Ultrasound		& Noise artifacts, median filtering, JPEG 2000 (3 ratios)		& Direct comparison of metrics; JAFROC: 5 radiologists (MRI) 	& NIQE, BiQES, NIQE-K (proposed modified NIQE)
        \tabularnewline
        
		\bottomrule
    \end{tabular}
    \label{tab:vq_based_nr2}
\end{table*}

\subsubsection{Endoscopic / Laparoscopic videos}

Based on a new publicly available database of 2D laparoscopic videos (LVQ)~\cite{khan2020towards}, a residual network-based method was proposed in \cite{khan2020residual} to predict the quality score and detect the type of distortion. Particularly, to tackle the problem of limited data, a ranking based pre-training approach has been proposed. The proposed model with a ResNet-18 backbone, obtained a better performance in terms of \ac{SROCC} with the established quality ranking (0.69), when compared to another deep learning-based method, i.e. RankIQA~\cite{rankIQA} (0.57). It should be noted that the quality ranking was inferred directly from the severity levels of simulated distortions (i.e., defocus and motion blur, Gaussian noise, uneven illumination, and smoke). Simultaneously, the authors attempted image classification into 20 different labels, considering every distortion-level pair. Different ResNet depths were tested, with ResNet-50 achieving the best accuracy (87.3\%).

In a very recent paper, Ali \textit{et al.}\cite{ali2021deep} proposed an integrated deep learning approach for both \ac{QA} and frame restoration in video endoscopy. In the context of this paper, the \ac{QA} method relied on detecting six different types of impairments, i.e., bubbles, blur, contrast, specularity, saturation, and miscellaneous artifacts, in near real-time, using a YOLOv3 architecture. The dataset included a set of frames from both normal bright field and narrow-band imaging videos of different patients. After a bounding box detection stage, the framework also performed a finer segmentation of the artifacts, with the best results being obtained with DeepLabv3+ spatial pyramid pooling.  

\subsubsection{Ultrasonography}

Abdi \textit{et al.}\cite{abdi2017automatic} trained a total of 430 CNN models to attempt the automated \ac{QA} of transthoracic echocardiograms, using \ac{PSO} for hyperparameter optimization. Subjective scoring was based on the visibility of several anatomical structures (\ac{MES}). The network resulting from \ac{PSO} featured three convolutional layers, fed to two fully connected layers, and was trained independently three times. The final model performance was measured by the mean absolute error between the predicted score and the \ac{MES}, with average reported values of 0.71 $\pm$ 0.58. The authors further analyzed the obtained CNN feature maps, which suggested the visibility of the septum and lateral walls to be important factors for a higher predicted image quality.

\subsubsection{Fused images}

Tang \textit{et al.} published two papers on the quality assessment of \ac{MMIF}. In both papers, the authors tested eight image fusion algorithms, namely, \ac{NNM}, \ac{LP-SR}, \ac{CSCS}, \ac{GFF} \ac{NSCT-PCNN-SF}, \ac{ISML}, \ac{CSR}, and \ac{DTM-PCNN}, with different pairs of imaging modalities, namely, \ac{CT} and \ac{MRI}, T1\emph{w} and T2\emph{w} MRI, B-mode Ultrasound and \ac{SPECT}, \ac{MRI} and PET, and \ac{MRI} and \ac{SPECT}. In \cite{tang2018no}, the quality index was obtained by pooling the phase congruency, which measures the correlation between salient features of the source and fused images, and the standard deviation, which provides a measure of sharpness and clarity. In ~\cite{TANG2020SPIC}, the authors further proposed using a PCNN with NSCT sub-images (both high- and low-frequency components). An overall quality index was then given by pooling the obtained components. The performance of the proposed metrics was compared to several state-of-the-art metrics, including gradient, structure, edge and entropy-based metrics, considering their correlation to the \ac{MOS} of twenty radiologists. In \cite{tang2018no}, the proposed metric achieved average a SROCC near 0.8 and KRCC near 0.7, whilst in the more recent paper, the proposed metric yielded the following performance measures: PLCC=0.79, SROCC=0.73, KRCC=0.61, and RMSE=0.27. Although both metrics largely outperformed the baseline metrics, suggesting their relevance in \ac{MMIF} \ac{QA}, the obtained correlations could be further improved.

\subsubsection{Multiple imaging modalities}

In \cite{outtas.study.16}, the authors proposed to modify \ac{NIQE}~\cite{niqe}, a \ac{NR} metric originally developed for natural images, and used for the \ac{QA} of medical images in some studies. The authors improved its perceptual evaluation, using a frequency-domain analysis inspired by the \ac{BIQES} metric~\cite{saha2015utilizing}. The ratio kurtosis/standard deviation of the log amplitude of the Fourier spectra was taken as weighting factor. In a first experiment, ultrasound images were corrupted with simulated noises, i.e. Sattar's noise and speckle noise, and then filtered using a median filter. NIQE-K and \ac{BIQES} achieved comparable performances in terms of quality ranking. Nonetheless, NIQE-K ranked a noisy image (Sattar's noise) better than the original ultrasound image, despite their close quality indices. In the second experiment, three radiologists were asked to the detect and localize multiple sclerosis lesions in \ac{MRI}. Their performance was quantified by the \ac{JAFROC}. Result analysis showed the consistency of NIQE-K for ultrasound image quality, and that, for 40\% of the images, the behavior of NIQE-K is the same as that of the radiologists.

\section{Task-based Methods}\label{TaskApp}

Task-based objective \ac{QA} methods, also often referred to as \ac{NO} or \ac{MO}, are designed to approach the performance of human observers (i.e., medical experts) in a given task, as opposed to visual quality-based approaches, which typically aim at predicting the \ac{MOS} of human observers~\cite{zhang2012perceptually}.

The underlying paradigm is to quantify the quality of medical images and videos by their effectiveness with respect to its intended purpose~\cite{zhang2012key}. In this section, we review task-based methods proposed for different tasks, as well as the corresponding \ac{FOM} for performance evaluation. Table \ref{tab:task_based_approach} is a summary of these works (note that we focus on works published after 2013, since an overview was already published then \cite{zhang2014overview}).

\subsection{Detection and classification tasks}
The most common task is the detection task, in which a decision is made in favor of one of two hypotheses, i.e., signal present (H1) or absent (H2). Low-dose reconstruction methods using iterative algorithms in tomography introduced new challenges related to extraction of image information. In any diagnosis, a good compromise must be found between the dose level and the resulting quality of the medical image, in order to reduce the exposure dose, while allowing the detection of low contrast textures. In \cite{ECK2015}, the authors used a \ac{CHO}~\cite{brankov2013evaluation, zhang2014overview} and developed an internal noise model to compare detectability indices ($d'$) in low-dose \ac{CT} images. The $d'$ index, chosen as \ac{FOM}, was calculated from the distribution of signal and noise decision variables (i.e., sum of channel outputs). The experiment was performed with five observers, using \ac{CT} phantom images, and showed that Iterative Model Reconstruction (IMR, Philips) enables at least a 67\% dose reduction comparatively to \ac{FBP}. 

Racine \textit{et al.}\cite{Racine2016ojective} conducted research to objectively evaluate the low contrast detectability in CT, with different radiation doses (CTDI$_{vol}$ of 5, 10, 15, and 20 mGy). Images of a QRM 401 phantom containing 5- and 8-mm diameter spheres, with a contrast level of 10 and 20 HU, were acquired and reconstructed using three algorithms, i.e., \ac{FBP}, \ac{ASIR}, and \ac{MBIR}. A \ac{CHO} model with \ac{D-DOG} channels was used to evaluate the image quality for every combination of the aforementioned parameters. The performance of the \ac{CHO} model was compared with that of six medical students, who provided detectability levels for the test images in four-alternative forced choice tests. A high correlation was found between the results of the human observers and the \ac{CHO} model, independently of the dose levels or the signals considered, suggesting it might be used to predict expected detectability levels, and ensure the diagnostic quality of low-dose \ac{CT} acquisitions. \ac{PLCC} was 0.98 for \ac{MBIR} and 0.93 for \ac{FBP}. \ac{MBIR} gave the highest overall detectability index, particularly for low CTDI$_{vol}$.

Greffier \textit{et al.}\cite{greffier2020ct} used the \textit{imQuest} software to assess the quality of low-dose \ac{CT} scans, comparing the performance of four manufacturers (i.e., GE, Philips, Siemens, and Canon), with different reconstruction algorithms (i.e., \ac{FBP}, \ac{MBIR} and \ac{H/SIR}), and five dose levels (i.e., 0.5, 1.5, 3.0, 7.0, and 12.0 mGy). A \ac{NPWMF} model observer with an eye filter~\cite{richard2008comparison} was used to calculate the detectability index, considering two detection tasks: a large mass in the liver, and a small calcification. The $d'$ index was obtained from the \ac{NPS} and the \ac{TTF}, which were measured on a ACR QA phantom with acrylic inserts. The reported results showed that the use of an optimization algorithm (either \ac{H/SIR} or \ac{MBIR}) improves $d'$ for low-dose acquisition, when compared to \ac{FBP}. When directly comparing \ac{H/SIR} and \ac{MBIR}, the second led to an increase in $d'$ (measured at 3 mGy), as well as potential dose reductions for Siemens, GE, and Philips systems, with features sizes. Potential dose reduction using Philips (IMR) reached 62\% and 78\% for the small and large feature, respectively.

Supervised learning-based approaches for implementing model observers have become possible substitutes for numerical observers, particularly through the use of \ac{CNN}. Li \textit{et al.}\cite{Li2021Assessing} evaluated different CNN-based denoising methods on simulated planar scintigraphy images. The images were simulated for \ac{SKE/BKS} lesion detection, with a Gaussian signal and a lumpy object as random background. Mixed Poisson and Gaussian noise was added to the simulated images, both signal absent and signal present, to produce the final dataset. Three denoising encoder-decoder networks were tested, i.e., a linear \ac{CNN} (without ReLU activation layers), a non-linear \ac{CNN} with \ac{MSE} loss, and a ResNet-based model with a perceptual loss function. Their performance was then assessed using several observers for the detection of the lesion signals, i.e., a Bayesian \ac{IO}~\cite{barrett1993model, CORMACK2005325}, a CNN-based observer, a \ac{HO}~\cite{barrett1993model, zhang2014overview}, a \ac{CHO}, a \ac{RHO}, and a \ac{NPWMF}. The authors performed \ac{ROC} analysis and further computed a detection efficiency measure, given by $e \equiv AUC_{denoised} / AUC_{noisy}$, observing that, while an increase in network depth improved \ac{SSIM} and \ac{RMSE} measures, the performance of the detection task generally dropped, thus suggesting that denoising methods might cause a loss of statistical information in the image. For the \ac{CNN}-based observer, \ac{HO} and \ac{RHO}, the performance decreased with the increase in network depth. It is also argued that using non-task-based loss functions to optimize CNN-based denoising models might play an important part in that loss of information.

\begin{table*}[t!]
\renewcommand{\arraystretch}{1.4}
\centering
\small
\caption{Overview of task-based approaches to the quality assessment of medical images and video.}
\vspace{0.2cm}
\begin{tabular}{>{\raggedright}m{0.09\textwidth} >{\raggedright}m{0.12\textwidth} >{\raggedright}m{0.22\textwidth} >{\raggedright}m{0.18\textwidth} >{\raggedright}m{0.12\textwidth} >{\raggedright}m{0.12\textwidth}}
\toprule
\textbf{Reference} & \textbf{Imaging modality} & \textbf{Quality issue} & \textbf{Model} & \textbf{Task} & \textbf{Figure of merit}
\tabularnewline
\midrule

\cite{ECK2015} & CT phantom & Reconstruction: IMR, FPB/ low dose radiation & CHO & Detection & Detectability (d')
\tabularnewline

\cite{Racine2016ojective} & CT phantom & Reconstruction: FBP, ASIR, MBIR/ low dose radiation & CHO (D-DOG) & Detection & Detectability (d')
\tabularnewline

\cite{kopp2018cnn} & CT phantom & Detection of lesions in CT with different X-ray exposures & SR-MO, CNN-MO, CHO (Gabor) & Detection & ROC / AUC
\tabularnewline

\cite{greffier2020ct} & CT phantom & Reconstruction: Asir, Asir-V, IMR, iDose SAFIRE, ADMIRE, AIDR3D, First, FPB, H/SIR and MBIR / low dose radiation & NPWMF (imQuest tool)
& Detection & Detectability (d')
\tabularnewline

\cite{Zhou2019} & Computer-simulated images & Approximation of the IO and the HO by supervised learning methods to evaluate image quality for detection tasks & CNN-IO, SLNN-IO & Detection & ROC / AUC
\tabularnewline


\cite{alnowami2018deep} & Mammography & Study the minimum detectable contrast in mammography scans & CNN-MO & Detection & Contrast threshold
\tabularnewline

\cite{Li2021Assessing} & Simulated Planar Scintigraphy images & Comparison of CNN-based denoising methods & IO, CNN-MO, HO, RHO, NPWMF & Detection & ROC / AUC \tabularnewline

\cite{wu2017fuiqa} & Ultrasound & Quality assessment of acquisition of fetal ultrasound scans & Custom CNN-based model & Localization & Multiple structure visibility
\tabularnewline

\cite{Zhou2020} & Computer-simulated images & Approximation of the IO and the HO by supervised learning methods to evaluate image quality for localization tasks & Analytical IO, Scanning HO, MCMC-IO, CNN-IO & Localization & LROC
\tabularnewline

\cite{Lorente2020} & Computer-simulated images & Approximation of a human observer in defect localization tasks & U-Net & Localization & Accuracy
\tabularnewline

\cite{welikala2016automated} & Retinal fundus \newline photography & Quality assessment of acquisition of retinal images & Measures from segmented areas with SVM or decision tree classifier & Segmentation & ROC / AUC
\tabularnewline

\cite{rodrigues2019quality} & MRI & Compare the performance of the segmentation model with several no-reference quality metrics & Texture-based features with AdaBoost classifier & Segmentation & Dice overlap coefficient
\tabularnewline

\cite{alais2020fast} & Retinal fundus \newline photography & Image quality in macular region & Custom CNN, U-Net & Segmentation, Localization & Macula visibility
\tabularnewline

\bottomrule
\end{tabular}
\label{tab:task_based_approach}
\end{table*}

The authors in \cite{kopp2018cnn} tested two antropomorphic model observers to detect liver lesions, one based on \textit{softmax} regression (SR-MO) and a CNN-MO. A phantom with contrast targets of different diameters was scanned on a \ac{CT} scanner, at different X-ray exposures, and the images were reconstructed using both \ac{FBP}, which was used for the test dataset, and iterative reconstruction, used in training and validation. One radiologist provided confidence levels considering the detection task for a total of 7488 images. Model performance evaluation relied on computing the \ac{JAFROC} for the reader study, a \ac{CHO} with Gabor channels, and the two proposed models, which were trained using two strategies, i.e., separated models for each lesion size, or a common model for all diameters. The \ac{PLCC}, the $\chi^2$ goodness-of-fit, and the \ac{MAPD} were then used to compare the models' \ac{AUC} values. The authors concluded that the CNN-MO can accurately approximate the human performance. This model outperformed the \ac{CHO} (\ac{PLCC} = 0.95, \ac{MAPD} = 2.2\%) using the first training strategy (\ac{PLCC} = 0.98, \ac{MAPD} = 1.2\%), and the SR-MO with both strategies. With the second strategy, the CNN-MO achieved a \ac{PLCC} of 0.92, and \ac{MAPD} of 3.0\%.

Alnowami \textit{et al.}\cite{alnowami2018deep} aimed at studying the minimum detectable contrast in mammography screenings, using a deep learning-based \ac{MO}. They first trained a data-driven \ac{CNN} model to classify image patches, from clinical routine screenings, as normal tissue or containing a lesion. In this SKS detection task, the model achieved a sensitivity of 0.90 and a specificity of 0.92 on the test set. In a second experiment, its performance was compared with two groups of human observers (experts and non-experts), using a four-alternative forced choice setup with simulated images to assess the minimum detectable contrast. Contrast threshold values obtained with the proposed \ac{MO} approximated the human performance on spherical targets, and even outperformed it for 4mm targets. For lesion targets, human performance was better, but the CNN-MO still achieved a comparable performance (12\% difference). 

Zhou \textit{et al.}\cite{Zhou2019} used deep learning methods to approximate an \ac{IO} and a \ac{HO} for binary signal detection tasks, i.e., a \ac{CNN} and a \ac{SLNN}, respectively. Computer-simulated images were generated using continuous-to-discrete mapping with a Gaussian kernel, and considering four binary signal detection tasks: a \ac{SKE/BKE} task, where the \ac{IO} and \ac{HO} are analytically determined, two \ac{SKE/BKS} tasks, with lumpy background and clustered lumpy background object models, and a \ac{SKS/BKS} task with a lumpy background. Overall, the CNN-IO and SLNN-HO closely approximated the results of the \ac{IO} and \ac{HO}, in terms of \ac{AUC}. In the case of the \ac{SKE/BKE} task, the obtained \ac{AUC} with \ac{IO} and CNN-IO was 0.89, whereas with \ac{HO} and SLNN-HO it was 0.83. For the \ac{SKE/BKS} task with lumpy background, the \ac{IO} was computed using \ac{MCMC} techniques~\cite{kupinski2003ideal}, and achieved an \ac{AUC} of 0.91. CNN-IO closely approximated this performance, with \ac{AUC}=0.91. The traditional \ac{HO} and SLNN-HO yielded an \ac{AUC} of 0.81. A similar outcome was observed in the \ac{SKE/BKS} task with clustered lumpy background for \ac{HO} and SLNN-HO (\ac{AUC}=0.85). CNN-IO achieved an \ac{AUC} of 0.89, but its performance could not be compared to the \ac{IO}, with the authors arguing that \ac{MCMC} application to clustered lumpy background object models had not been reported to date. Finally, for the \ac{SKS/BKS} task, the performance of both the traditional \ac{HO} and SLNN-HO are close to a random estimate (\ac{AUC} $\approx$ 0.5), as linear observers are generally unable to detect signals with random locations. As for MCMC-IO and CNN-IO, the performance was again very close (\ac{AUC}=0.86).

\subsection{Localization tasks}

An interesting work was published in \cite{wu2017fuiqa}, where the authors proposed a localization-based approach to assess the quality of acquisition of fetal ultrasound scans. First, the authors implemented a localization CNN (L-CNN) using pre-learned AlexNet low-level cues, which identifies a \ac{ROI} containing the fetal abdomen. A sliding window strategy was used to feed local inputs to the L-CNN and obtain a \ac{ROI} probability map throughout the entire scan. Image quality was then assessed by considering the quality of depiction of both the \ac{SB} and \ac{UV} within the \ac{ROI}. These two structures may be labeled as satisfactory, not good or absent. While the first label translates to $S_{SB}$ or $S_{UV}$=1, the other two led to 0. From the combinations of these two binary scores, a 4-class output was obtained by either human observers or a classification CNN (C-CNN), with knowledge transferred from the L-CNN to its encoding layers. The final quality score of the \ac{FUIQA} scheme is given by the sum of $S_{SB}$, $S_{UV}$ and $S_{ROI}$, which is also a binary score, determined by the \ac{ROI} to field of view ratio. The authors provided an extensive discussion on the proposed methods, for example comparing the \ac{AUC} for \ac{SB} and \ac{UV} detection with different input layers, based on local phase analysis. The reported quality outputs of the described \ac{FUIQA} model were highly coherent with those from 3 radiologists, with agreement values of 0.91, 0.89, and 0.88.

Zhou \textit{et al.} \cite{Zhou2020} extended their proposed supervised learning method~\cite{Zhou2019} to approximate an \ac{IO} in joint detection and localization tasks. The LROC curves produced by the proposed method were compared with those of traditional observers when computationally feasible. The signal can be localized in nine different locations and the considered signal detection-localization tasks are BKE task, BKS with a lumpy background model and BKS with a clustered lumpy background model. Overall, the CNN-IO was able to come close to the analytical IO for the BKE task, the MCMC-IO for the BKS task with a lumpy background, and outperformed the scanning HO for the BKS task with CLB since MCMCs have not been reported to date for this type of background. Note that approximating an IO using supervised learning requires a large amount of training data, which can be a challenge when only a limited amount of experimental data is available.

Lorente \textit{et al.}\cite{Lorente2020} proposed a \ac{MO} based on U-Net~\cite{unet} for defect localization on simulated images, with three levels of correlated noisy backgrounds. Two network configurations (i.e., kernel sizes 3×3 and 5×5) and two loss functions (i.e., \ac{MSE} and binary cross-entropy) were tested, and their accuracy compared with a human observer. Accuracy was described as the ratio between correctly located defects, i.e. within a 5 pixel distance of the actual defect, and the total number of images. The authors concluded that the models trained with a binary cross-entropy loss function provided results closer to the human observer. For example, for the third level of correlated noisy background, the accuracy of the human observer was 0.8, and the 3x3 and 5x5 \ac{MO} trained with \ac{MSE} loss yielded an accuracy of 0.92 and 0.91, respectively. The \ac{MO} trained with binary cross-entropy loss, got accuracy values of 0.89 and 0.85, respectively.

\subsection{Segmentation tasks}

In \cite{welikala2016automated}, the authors studied a segmentation-based \ac{QA} framework for retinal images. Unsupervised vessel segmentation was performed on a dataset of 800 images from the UK Biobank~\cite{sudlow2015uk}, using QUARTZ (Quantitative Analysis of Retinal Vessel Topology and Size)~\cite{quartz}. The algorithm relies on a multi-scale line detector, based on the average gray-level intensities around each target pixel, complemented with high-intensity pixel thresholding, masking of the fovea, and suppression of small objects. Image quality was then measured by extracting three features from the binary segmentation image, selected to mimic manual \ac{QA}, i.e., area, fragmentation, and complexity, which were finally classified using both a \ac{SVM} classifier with a radial basis function kernel and an ensemble decision tree classifier. The dataset had been labeled by two observers as adequate or inadequate for epidemiological studies. The \ac{SVM} achieved the best performance in classification, with an \ac{AUC} of 0.98. Although method validation is not directly related with the human segmentation performance, the method uses the segmentation task outcome to predict image quality.

The authors in \cite{rodrigues2019quality} studied the correlation between texture-based muscle segmentation in Dixon \ac{MRI} and a set of \ac{NR} quality metrics, i.e., Variance, Laplacian, Gradient, Autocorrelation, \ac{FTM}, \ac{MarzBM}, HP metric, Kurtosis-based metric, and \ac{RTBM}, to assess the feasibility of using texture segmentation methods as a content-specific quality measure for \ac{MRI}. The authors implemented a pixel-wise binary segmentation method, using AdaBoost~\cite{Adaboost}, with a local texture descriptor, consisting of the histogram of oriented gradients (HOG)~\cite{HOG}, 3-level Haar Wavelet coefficients, and statistical measures from the original grayscale image and the Laplacian of Gaussian (LoG) filter~\cite{LOG}. The Dice overlap coefficient with manual segmentations was chosen as \ac{FOM}. Overall, the segmentation output showed reasonable correlation with the variance metric (\ac{PLCC} = 0.72, \ac{SROCC} = 0.74) and \ac{RTBM} (\ac{PLCC} = 0.71, \ac{SROCC} = 0.73). Considering only cases with poor segmentation (Dice $<$ 0.7), where all metrics performed better, even though it should be noted that this analysis relied on a smaller number of points. \ac{RTBM} (\ac{PLCC} = 0.93, \ac{SROCC} = 0.86) and FTM (\ac{PLCC} = 0.84, \ac{SROCC} = 0.96) yielded the best correlations.

More recently, Alais \textit{et al.}\cite{alais2020fast} proposed a \ac{CNN} that segments the macular region, which consists of 3×3 convolutional layers only, has very few parameters, and makes decisions considering a threshold $t$ to obtain a binary image. If the obtained area is greater than a value $A$, then the algorithm considers that the macula is visible, and the image quality is considered sufficient for macula location. The authors extracted 6098 eye fundus images from the \textit{e-ophtha} database~\cite{decenciere2013teleophta}, with more than half of the images containing a visible macula. The chosen parameters, $t$ and $A$, that offer a trade-off between sensitivity and specificity, gave an accuracy of 96.4\% on the test set. The proposed \ac{CNN} wrongly predicted that the macula was visible in four images among 304 images (i.e., 1.3\% of false positives), whereas U-Net had nine false positives. Regarding the fovea localization, the authors obtained an average error of 0.95 pixels for their network versus 1.22 pixels for U-Net. The same tests, conducted on the ARIA database~\cite{farnell2008enhancement}, revealed a prediction of the fovea with 1.4 pixels of mean error (0.1mm) and 6 pixels of maximum error. Finally, the network was tested on pathological images, with one unsuccessful detection due to the presence of macular hemorrhages or exudates.

\section{Discussion and Future Work}

Medical images and video refer to a wide variety of different acquisition methods, clinical applications, and quality issues, as it is clear from the papers reviewed in this paper. Therefore, it is almost impossible to establish meaningful comparisons between the objective methodologies used, i.e., their merits and drawbacks. Nonetheless, some global considerations may be drawn, particularly on future research directions.

Regarding \ac{FR} and \ac{RR} visual quality-based metrics, all reviewed papers reported the use of metrics that were originally designed for natural content, or variations of those metrics. \ac{PSNR} and \ac{SSIM} were the most common across these studies, and both were used in a vast majority of them~\cite{chow2016correlation, mason2019comparison, zhou2003practice, razaak2014study, razaak2016cuqi, chaabouni14, usman2017quality, kumar2009mos, kumar2013development, nasr2017visual}. \cite{renieblas2017structural} also reported the use of \ac{SSIM}, along with several variations of that metric. Other metrics, such as \ac{VIF} and \ac{NQM} were commonly tested as well. The only \ac{FR} metric specifically designed for medical imaging was proposed by \cite{razaak2016cuqi}, who described a \ac{QA} metric for cardiac ultrasound videos, based on cardiac motion and structural information. Considering the reviewed \ac{NR} visual quality-based approaches, most methods are tailored for specific imaging modalities and/or artifacts. From our analysis, it is also clear that deep learning methods are becoming a staple in \ac{NR} medical image and video \ac{QA}, as most recent papers used \ac{CNN} instead of handcrafted features, with some interesting results~\cite{abdi2017automatic, esses2018automated, ma2020diagnostic, coyner2019automated, sujit2019automated, raj2020multivariate, shen2020domain, khan2020residual, ali2021deep}.

As stated in \cite{leveque2021}, there seems to be a lack of publicly available databases with subjective quality annotations in the medical imaging domain. Some notable examples are described in \cite{Suad2013, di2014autism, Outtas2018, khan2020towards}. In order to develop new reliable objective quality metrics for the medical imaging field, more training data is needed. With a view to address this issue, human-in-the-loop machine learning techniques could be considered (e.g., as described by \cite{Willemink2020}). In this sense, although automated and semi-automated techniques have been proposed for segmentation purposes~\cite{welikala2016automated, rodrigues2019quality, alais2020fast}, there is still a lack of annotated databases and studies to support the development of reliable methods, for example incorporating models of how clinicians perform diagnosis from images and videos~\cite{Alexander2020}. Artificial intelligence promises strong breakthrough in medical imaging objective \ac{QA}.

Another common challenge in objective \ac{QA} studies for medical images and videos has to do with artifact simulation. While some of the reported visual quality-based studies used databases containing images or video with real artifacts (e.g., \cite{lalonde2001automatic, wang2015human, Liebgott16ICASSP, esses2018automated, sujit2019automated, ma2020diagnostic, shen2020domain, ali2021deep}), others used simulated distortions and/or artifacts over real data assumed to be clean (e.g., \cite{nasr2017visual, mason2019comparison, chow2016correlation, renieblas2017structural, osadebey2017bayesian, chabert2021image, kohler2013automatic, outtas.study.16, khan2020residual}). Collecting data with real artifacts may not always be possible, or ends up being impractical. On the other hand, simulated artifacts are often generic and limited in range, which might hinder the application of the developed \ac{QA} methods to real clinical data~\cite{oh2021unpaired}. Recently, some efforts towards the simulation of content-specific and realistic artifacts, to be applied in healthcare \ac{QA} research, have been reported (e.g., \cite{yang2019robust, oktaviana2019preliminary, hu2021simulation, oh2021unpaired}), and future work will likely approach this question more often, leveraging on the continuous development of deep-learning methods, such as \ac{GAN}.

A similar question can be raised regarding task-based \ac{QA} research. Traditional model observers are based on the (total or partial) knowledge of statistical characteristics of the images (i.e., signal and background). Hence, many studies (e.g.,~\cite{kalayeh2013, ECK2015, greffier2020ct, Racine2016ojective, Zhou2019, Zhou2020, Lorente2020}) are based on simulated or phantom images, since real medical images suffer from a lack of statistical information. Currently, there is no evidence that \ac{QA} studies conducted on simulated images ensure sufficient confidence to draw relevant conclusions on real images. Deep learning methods could allow to go through this weakness. Indeed, any task that could be carried out with deep learning provides an assessment of the quality. In fact, the more efficiently the task is done, the better the quality. There is still room for improvement in the field, the whole purpose being to define the most relevant tasks, i.e., those that can be reliably delegated to a model. In task-based \ac{QA}, the key problem lies in modeling the task performed by human observers. So far, to our knowledge, existing models are limited in task range. For example, no model observer has been proposed for characterization tasks, which focus on analyzing certain properties of abnormalities (e.g., contour or texture) for differential diagnosis, and generally involve a linguistic response (e.g., benign vs. malignant), given its high complexity~\cite{zhang2014overview}. Other tasks could also be further explored in future work, such as estimation tasks (or joint detection / classification / estimation tasks), which aim at determining a scalar value or range of values from a given object to be used in diagnosis (e.g., tumor diameter or radiotracer uptake~\cite{barrett2015task}).

Although 3D visualization of medical images and videos is emerging, research on relevant quality assessment aspects is still behind. Stereoscopic medical imaging and, more recently, light field medical imaging, open new opportunities, for instance in surgery training, also at a distance~\cite{martini20133}. Compression~\cite{nagoor2020lossless} and transmission~\cite{martini20133} of 3D stereoscopic medical images and videos, as well as of light field medical data, require suitable metrics for the assessment of their performance.
Recently, studies on quality assessment for light field medical images have started (e.g, \cite{kara2017perceptual}). However, objective metrics in this domain are still missing, or their development is still ongoing.
Future research might focus on assessing whether existing metrics, developed for generic 3D images and videos (e.g.,
\cite{han2016innovative},
\cite{hewage2013quality},
\cite{hewage2011reduced},
\cite{battisti2015objective}),
and light field data
(e.g., \cite{ak2019investigating},
\cite{tamboli20183d},
\cite{tamboli2018objective},
\cite{viola2016objective}), are
suitable to assess the quality of medical data represented in these formats. While efforts are ongoing in this direction, the availability of wider datasets of medical data in stereoscopic 3D and light field 3D formats would definitely be useful towards this effort.



Several of the reported studies considered coding distortions \cite{chow2016correlation, mason2019comparison, zhou2003practice, kowalik2015modelling, panayides2011atherosclerotic, razaak2014study, razaak2016cuqi, kumcu2014visual, chaabouni14, usman2017quality, kumar2009mos, kumar2013development, renieblas2017structural, planitz2005study, nasr2017visual, chow2017modified, outtas.study.16}. There is a number of applications, mostly based on telemedicine applications, where lossy compression of medical images or videos might be acceptable. However, in most cases, medical imaging applications cannot rely on images with lossy compression, as no one can be sure of the influence that those losses can have in a diagnosis. Multiple times, radiologists use almost imperceptible textures to define their diagnosis and, in such cases, lossy compression can have a major impact. Hence, several studies on medical image quality are quite questionable, as they use medical modalities where no radiologist would accept any kind of lossy coding.

\section{Conclusion}

In this article, we presented an extensive review of the literature on the objective quality assessment of medical images and video. More precisely, we covered both visual quality-based and task-based methods, considering various imaging modalities and application purposes. Furthermore, we aimed at covering a wide range of approaches to the quality assessment of medical images and videos content, including the application of preexisting metrics for natural images and the development of content-specific metrics, either based on handcrafted features or using deep learning-based models.

\section*{Acknowledgements}
R. Rodrigues and A. Pinheiro acknowledge Funda\c{c}\~ao para a Ci\^encia e a Tecnologia (FCT/MCTES) for funding this research, under the doctoral grant SFRH/BD/130858/2017, and the project UIDB/50008/2020, respectively. They would also like to acknowledge the project CENTRO-01-0145-FEDER-000019 - C4 Cloud Computing Competence Centre. The work of J. Gutiérrez was partially supported by the Juan de la Cierva fellowship (IJC2018-037816).

\bibliographystyle{spphys}
\bibliography{refsBibTex}

\begin{acronym}
\acro{CT}{Computed Tomography}
\acro{MRI}{Magnetic Resonance Imaging}
\acro{PET}{Positron Emission Tomography}
\acro{SPECT}{Single-photon Emission Computed Tomography}
\acro{OECD}{Organization for Economic Co-operation and Development}
\acro{QoE}{Quality of Experience}
\acro{QA}{quality assessment}
\acro{MOS}{Mean Opinion Score}
\acro{DMOS}{Differential MOS}
\acro{FR}{Full Reference}
\acro{NR}{No Reference}
\acro{RR}{Reduced Reference}
\acro{RMSE}{Root Mean Square Error}
\acro{OR}{Outlier Ratio}
\acro{DSCQS}{Double Stimulus Continuous Quality Scale}
\acro{SSCQE}{Single Stimulus Continuous Quality Evaluation}
\acro{PSNR}{Peak Signal-to-noise Ratio}
\acro{MSE}{Mean Squared Error}
\acro{SSIM}{Structural Similarity index}
\acro{MS-SSIM}{Multi-scale SSIM}
\acro{UQI}{Universal Quality Index}
\acro{IFC}{Information Fidelity Criterion}
\acro{VIF}{Visual Information Fidelity}
\acro{VIFP}{Pixel Visual Information Fidelity}
\acro{BRISQUE}{Blind/Referenceless Image Spatial Quality Elevator}
\acro{PLCC}{Pearson Linear Correlation Coefficient}
\acro{SROCC}{Spearman Rank-order Correlation Coefficient}
\acro{VQM}{Video Quality Metric}
\acro{KRCC}{Kendall Rank Correlation Coefficient}
\acro{FSIM}{Feature Similarity Index}
\acro{NQM}{Noise Quality Measure}
\acro{VSNR}{Visual Signal-to-noise Ratio}
\acro{SNR}{Signal-to-noise Ratio}
\acro{ROC}{Receiver Operating Characteristic}
\acro{AUC}{Area Under the Curve}
\acro{SDSCE}{Simultaneous Double Stimulus for Continuous Evaluation}
\acro{HEVC}{High Efficiency Video Coding}
\acro{MMIF}{Multimodal Medical Image Fusion}
\acro{DCT}{Discrete Cosine Transform}
\acro{SPIHT}{Set Partitioning in Hierarchical Trees}
\acro{WSNR}{Weighted Signal-to-noise Ratio}
\acro{SVR}{Support Vector Regression}
\acro{FLAIR}{Fluid Attenuated Inversion Recovery}
\acro{ANOVA}{Analysis of Variance}
\acro{AS-OCT}{Anterior Segment Optical Coherence Tomography}
\acro{LBP}{Local Binary Patterns}
\acro{PCA}{Principal Component Analysis}
\acro{SVM}{Support Vector Machine}
\acro{CRVM}{Channelized Relevance Vector Machine}
\acro{CSVM}{Channelized SVM}
\acro{MKCRVM}{Multi-Kernel CRVM}
\acro{CHO}{Channelized Hotelling Observer}
\acro{CNN}{Convolutional Neural Networks}
\acro{NIQE}{Natural Image Quality Evaluator}
\acro{HDR-VDP-2}{High Dynamic Range Visible Difference Predictor v2}
\acro{HVS}{Human Visual System}
\acro{GLCM}{Gray-level Co-occurrence Matrix}
\acro{MSSIM}{Mean SSIM}
\acro{CUQI}{Cardiac Ultrasound Video Quality Index}
\acro{CNR}{Contrast-to-noise Ratio}
\acro{FBER}{Foreground-Background Energy Ratio}
\acro{ROI}{Regions of Interest}
\acro{LDA}{Linear Discriminant Analysis}
\acro{QDA}{Quadratic Discriminant Analysis}
\acro{MLP}{Multilayer Perceptron}
\acro{CPBD}{Cumulative Probability of Blur Detection}
\acro{BIQES}{Blind Image Quality Evaluator based on Scales}
\acro{BPRI}{Blind Pseudo-reference Image-based metric}
\acro{IL-NIQE}{Integrated Local NIQE}
\acro{QENI}{Quality Evaluator of Noisy Images}
\acro{SISBLIM}{Six-step Blind Metric}
\acro{SSEQ}{Spatial–Spectral Entropy-based Quality}
\acro{MEON}{Multi-task End-to-end Optimized Deep Neural Network}
\acro{dipIQ}{Quality-discriminable Image Pairs Inferred Quality}
\acro{SNR-ROI}{Region-of-Interest SNR}
\acro{PSO}{Particle Swarm Optimization}
\acro{MES}{Manual Echo Score}
\acro{FBP}{Filtered Back-projection}
\acro{MBIR}{Model-based Iterative Reconstruction}
\acro{ASIR}{Adaptive Statistical Iterative Reconstruction}
\acro{D-DOG}{Dense Difference-of-Gaussian}
\acro{FOM}{Figures of Merit}
\acro{IR}{Iterative Reconstruction}
\acro{NPWMF}{Non-Prewhitening Matched Filter}
\acro{H/SIR}{Hybrid/Statistical Iterative reconstruction}
\acro{NPS}{Noise Power Spectrum}
\acro{TTF}{Task Transfer Function}
\acro{IO}{Ideal Observer}
\acro{HO}{Hotelling Observer}
\acro{SLNN}{Single Layer Neural Network}
\acro{NNM}{Nuclear Norm Minimization}
\acro{LP-SR}{Laplacian Pyramid and Sparse Representation}
\acro{CSCS}{Cross-scale Coefficient Selection}
\acro{GFF}{Guided Filtering}
\acro{NSCT-PCNN-SF}{Pulse-coupled Neural Network with Modified Spatial Frequency based on Non-subsampled Contourlet Transform}
\acro{ISML}{Improved Sum-Modified-Laplacian}
\acro{CSR}{Convolutional Sparse Representation}
\acro{DTM-PCNN}{Discrete Tchebichef Moments and Pulse-coupled Neural Network}
\acro{FTM}{Frequency Threshold metric}
\acro{MarzBM}{Marziliano Blurring metric}
\acro{RTBM}{Riemannian Tensor-based metric}
\acro{SKE/BKE}{Signal Known Exactly / Background Known Exactly}
\acro{SKE/BKS}{Signal Known Exactly / Background Known Statistically}
\acro{SKS/BKS}{Signal Known Statistically / Background Known Statistically}
\acro{MCMC}{Markov-chain Monte Carlo}
\acro{JAFROC}{Jack-knife Alternative Free-response Receiver Operating Characteristic}
\acro{MAPD}{mean absolute percentage difference}
\acro{NO}{\textit{numerical observers}}
\acro{MO}{\textit{model observers}}
\acro{FUIQA}{Fetal Ultrasound Image Quality Assessment}
\acro{DCNN}{Deep Convolutional Neural Network}
\acro{SB}{stomach bubble}
\acro{UV}{umbilical vein}
\acro{RHO}{Regularized HO}
\acro{ADDA}{Adversarial Discriminative Domain Adaptation}
\acro{DRIQC}{Diabetic Retinopathy Image Quality Classification}
\acro{MFIQA}{Multi-task Fundus Image Quality Assessment}
\acro{GAN}{Generative Adversarial Networks}
\acro{GMSD}{Gradient Magnitude Similarity Deviation}
\acro{JNB}{Just Noticeable Blur}

\end{acronym}

\end{document}